\documentclass[aps,prl,twocolumn,superscriptaddress,showpacs]{revtex4}

\usepackage{graphicx}
\usepackage{color}
\begin{document}

\author{Taisuke Ohta}
\affiliation{Advanced Light Source, Lawrence Berkeley National Laboratory, Berkeley, California, USA}
\affiliation{Fritz-Haber-Institut der Max-Planck-Gesellschaft, Berlin, Germany}

\author{Aaron Bostwick}
\affiliation{Advanced Light Source, Lawrence Berkeley National Laboratory, Berkeley, California, USA}

\author{J.\ L.\ McChesney}
\affiliation{Advanced Light Source, Lawrence Berkeley National Laboratory, Berkeley, California, USA}
\affiliation{Montana State University, Bozeman, Montana, USA}

\author{Thomas Seyller}
\affiliation{Institut f\"{u}r Physik der Kondensierten Materie, Universit\"{a}t Erlangen-N\"{u}rnberg, Erlangen, Germany}

\author{Karsten Horn}
\affiliation{Fritz-Haber-Institut der Max-Planck-Gesellschaft, Berlin, Germany}

\author{Eli Rotenberg}
\affiliation{Advanced Light Source, Lawrence Berkeley National Laboratory, Berkeley, California, USA}

\title{Interlayer interaction and electronic screening in multilayer graphene}
\date{\today}

\def\EF{$E_{\mathrm{F}}$}
\def\ED{$E_{\mathrm{D}}$}
\def\kpar{$k_{\parallel}$}
\def\kparnum{\kpar$=-1.703~$\AA$^{-1}$}
\def\kperp{$k_{\perp}$}
\def\red{\textcolor{red}}
\def\blue{\textcolor{blue}}
\def\PI{$\pi$}
\def\bthe{\blue{the}}
\def\gone{$\gamma_{1}$}
\def\h{$\cal H$}

\begin{abstract}
The unusual transport properties of graphene are the direct consequence of a peculiar bandstructure near the Dirac point. We
determine the shape of the \PI\ bands and their characteristic splitting, and the transition from a pure 2D to quasi-2D behavior
for 1 to 4 layers of graphene by angle-resolved photoemission. By exploiting the sensitivity of the \PI\ bands to the electronic
potential, we derive the layer-dependent carrier concentration, screening length and strength of interlayer interaction by
comparison with tight binding calculations, yielding a comprehensive description of multilayer graphene's electronic structure.
\end{abstract}
\maketitle

Much recent attention has been given to the electronic structure of multilayer films of graphene, the honeycomb carbon sheet which
is the building block of graphite, carbon nanotubes, C$_{60}$, and other mesoscopic forms of carbon \cite{Sai98}.  Recent progress
in synthesizing or isolating multilayer graphene films \cite{Nov04, Zha05, Ber04} has enabled access to their physical properties,
and revealed many interesting transport phenomena, including an anomalous quantum Hall effect \cite{Nov05, Zha05b}, ballistic
electron transport at room temperature \cite{Ber06}, micron-scale coherence length \cite{Mor05, Ber06} and novel many-body
couplings\cite{Bos06}.  These effects originate from the effectively massless Dirac Fermion character of the carriers derived from graphene's
valence bands, which exhibit a linear dispersion degenerate near the so-called Dirac point energy, \ED\ \cite{Wal47}.

These unconventional properties of graphene offer a new route to room temperature, molecular-scale electronics capable of quantum
computing \cite{Zha05b}.  For example, a possible switching function in bilayer graphene has been suggested by reversibly lifting
the band degeneracy at the Fermi level (\EF) upon application of an electric field \cite{Oht06, McC06}.  This effect is due to a
unique sensitivity of the bandstructure to the charge distribution brought about by the interplay between strong interlayer hopping
and weak interlayer screening, neither of which are currently well-understood.  

In order to evaluate the charge distribution, screening, stacking order and interlayer coupling, we systematically
studied the evolution of the bandstructure of one to four layers of graphene using angle-resolved photoemission spectroscopy
(ARPES).  We demonstrate experimentally that the interlayer interaction and stacking sequence affect the topology of the \PI\
bands, the former inducing an electronic transition from pure-2D to quasi-2D when going from one layer to multilayer graphene.  The interlayer hopping integral, screening length, and layer-dependent carrier concentration are determined as a function
of the number of graphene layers by exploiting the sensitivity of \PI\ states to the electronic potential. 

\begin{figure*}[htb]
%\begin{center}
    \includegraphics[width=11.5cm]{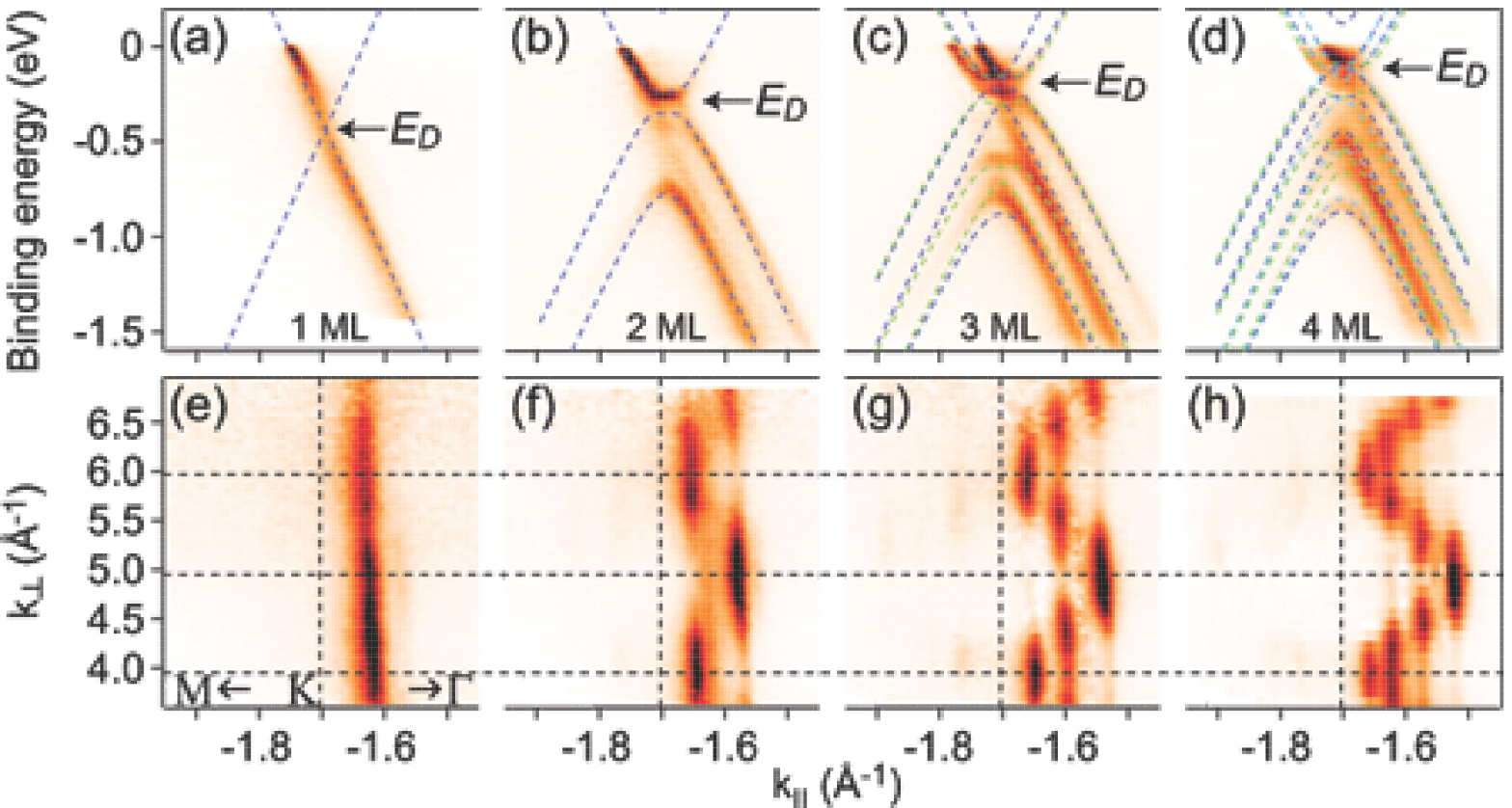}
% f02003_40048_92
% \vspace{1mm}
\caption{\label{fig:piBS} (color online) (a-d) The \PI\ and \PI* bands near \EF\ for 1 to 4 graphene layers, respectively.  \kparnum\ corresponds
to the K point, the corner of the hexagonal Brillouin zone.  The $\Gamma$ point is at \kpar\ =0 ~\AA$^{-1}$, while the M point is
at -2.555~\AA$^{-1}$.  The dashed lines are from a calculated tight binding bandstructure, with band parameters adjusted to
reproduce measured bands.  Dark and light blue lines are for Bernal-type (ABAB and ABAC) stackings, while green lines are for rhombohedral-type
stackings.  (e-h) Photoemission intensity oscillation of \PI\ bands at \EF $-1$ eV as a function of \kpar\ and \kperp\ momentum for
1-4 layers graphene.  The photoemission intensity is normalized by the angle integrated intensity between \EF\ and \EF$-1.5$ eV for
each photon energy.  }
%\end{center}
\end{figure*}

The films were synthesized on $n$-type (nitrogen, $1\times10^{18}\ \mathrm{cm}^{-3}$) 6H-SiC(0001) substrates (SiCrystal AG) that
were etched in hydrogen at 1550 C.  Annealing in a vacuum first removes the resulting silicate adlayer and then causes the growth
of the graphene layers between 1300 to 1400C\cite{For98}.  Beyond the first layer, the samples have a $\sim 1$ monolayer thickness
distribution; the bandstructures of different thicknesses were extracted using the method of Ref. \cite{Oht06}.  ARPES measurements
were conducted at the Electronic Structure Factory endstation at beamline 7.01 of the Advanced Light Source, equipped with a
Scienta R4000 electron energy analyzer.  The samples were cooled to $\sim 30$K by liquid He.  The photon energy was 94 eV with the
overall energy resolution of $\sim$30 meV for Fig.\ \ref{fig:piBS}(a-d).

\begin{figure}
\begin{center}
    \includegraphics[width=8.4cm]{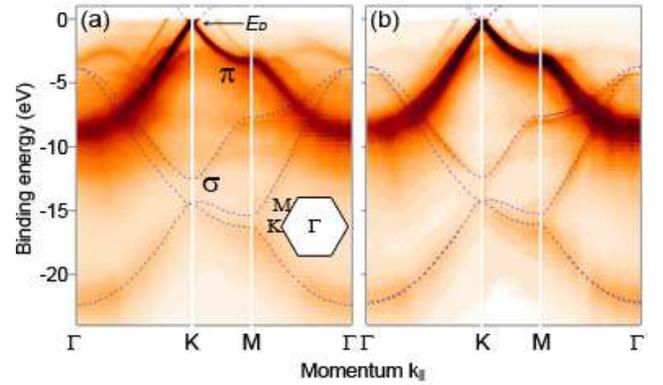}
% f01096band f01105band
% \vspace{1mm}
\caption{\label{fig:overallBS} (color online) Photoemission images revealing the bandstructure of (a) single and (b) bilayer graphene along high
symmetry directions, $\Gamma$-K-M-$\Gamma$.  The blue dashed lines are DFT bandstructure of free standing films \cite{Lat06}.
Inset in (a) shows the 2D Brillouin zone of graphene. }
\end{center}
\end{figure}

The bandstructures of a single (Fig.\ \ref{fig:overallBS} (a)) and a bilayer (b) of graphene are reflected in their photoemission intensity patterns as a
function of \kpar.  The data are compared with the scaled DFT bandstructures (see below) of free
standing graphene layers \cite{Lat06}.  We identify one \PI\ and three $\sigma$ bands, with the \PI\ band becoming split in the
bilayer.  At the K point, the constant energy contour for the \PI\ band becomes point-like near \EF\ \cite{Bos06, Oht06, Rol06},
similar to graphite at the bulk H point \cite{Zho06, Sugawara06}.  Other features represent either underlying substrate states
(especially in Fig.\ \ref{fig:overallBS}(a)) or replica bands due to diffraction from the underlying SiC surface reconstruction
\cite{Bos06,Oht06,Emt06}.

The DFT bands are shifted to account for the Fermi level position, and expanded by 13$\%$ in energy to match the experimental
total band width.  This scaling effectively incorporates many-body interactions not included in the theory, as shown in an earlier
ARPES study of graphite \cite{Hes99}.  The experimental bandwidths of the films are in close agreement with those of graphite
\cite{Hes99} and thick graphene multilayers \cite{Emt06}.

The \PI\ bands near \ED\ exhibit a complex structure due to interlayer interactions \cite{Oht06}.  Detailed photoemission intensity
maps near \EF\ are shown in Fig.\ \ref{fig:piBS}(a-d) for $N$=1 to 4 graphene layers.  The measured \PI\ bands are suppressed on
one side of the Brillouin zone due to interference effects between the two equivalent sublattices \cite{Shi95}.  The number of \PI\
bands increases with the number of layers due to interlayer splitting, clearly seen away from \ED, where the
Coulomb potential of each layer does not play a major role (see below).  The splitting between the highest and lowest \PI\ bands
increases with the number of layers and for quadlayer graphene (Fig.\ \ref{fig:piBS}(d)) it is close to that of bulk graphite,
$\sim$0.7 eV \cite{Cha91, Zho06}.  There is a gap between the \PI\ and \PI* bands in the bilayer (Fig.\ \ref{fig:piBS}b) due to the
inequivalent on-site Coulomb potentials in each layer\cite{Oht06}.

The \PI\ bands may be modeled by a tight binding (TB) calculation that takes
into account the different stacking sequences and on-site potential energies, with a Hamiltonian generalized from Refs.\ \cite{Gui06,McC06b, McC06}
as

\def\tr{\mathrm{T}}
\def\ai{\alpha_{i}=\left( \begin{array}{cc} E_{i} & v\pi^{\dag} \\ v\pi & E_{i} \end{array} \right)}
\def\bi{\beta_{s}=\gamma_{1}\left( \begin{array}{cc} 0 & s \\ 1-s & 0 \end{array} \right)}
\[
\cal H = \left( \begin{array}{cccccc}
\alpha_{1}      & \beta_{0}        &                 &                     &            &                    \\
\beta_{0}^{\tr} & \alpha_{2}       & \beta_{s}       &                     &            &                    \\
                & \beta_{s}^{\tr}  & \alpha_{3}      & \beta_{0}           &            &                    \\
                &                  & \beta_{0}^{\tr} & \alpha_{4}          & \beta_{s}  &                    \\
		&                  &                 & \beta_{s}^{\tr}     & \ddots     &                    \\
		&                  &                 &                     &            &\alpha_{\mathrm{N}} \\
\end{array} \right), 
\begin{array}{r}
    \ai \\
    \   \\
    \bi  \\
\end{array}.
\]
Here $E_{i}$ is the on-site Coulomb energy for layer $i$, $\pi=p_{x}+ip_{y}$, \gone\ is the interlayer hopping integral, $v$ is the
band velocity, and $s$=0 for Bernal (ABA\ldots) and 1 for rhombohedral (ABC\ldots) stacking.  The Hamiltonian operates on the layer
subspace $i=(1, 2, \ldots N)$ while the $2\times 2$ operators $\alpha$ and $\beta$ act on the $(A,B)$ sublattice sites of the same
or adjacent layers, respectively.  The energy scale was defined such that \EF=0 and the Dirac crossing energy \ED=Tr \h$/2N$.  The
Hamiltonian can be readily generalized to arbitrary stacking orders (e.g. ABAC) by suitable rearrangements of the coupling operators
$\beta_{s}$.

It is well known that ABA and ABC stackings for graphene are energetically close, and stacking faults are commonly found in highly
ordered pyrolytic graphite \cite{Moos01}.  For samples with mixed stacking, we assumed the same on-site Coulomb potential for all
stacking sequences.  For trilayer graphene (Fig.\ \ref{fig:piBS}(c)), we find two sets of \PI\ bands resulting from different
stacking sequences.  The blue and green dashed lines indicate TB bands with Bernal and rhombohedral stackings, respectively.  The
effect of different stackings is most apparent for the middle \PI\ band near \EF$-0.5$ eV, but is also seen in the upper and lower
\PI\ bands.

The situation is somewhat different for the quadlayer shown in Fig.\ \ref{fig:piBS}(d), where the measured and TB \PI\ bands are
compared.  The dark and light blue dashed lines are for Bernal-type stackings, ABAB and ABAC, respectively.  The green dashed lines
are for rhombohedral-type stackings, ABCA and ABCB, which are equivalent within the nearest neighbor TB model employed here.  We
assumed that the on-site Coulomb potentials between layers have the same sign and change monotonically.  Four \PI\ bands are
well-reproduced by assuming Bernal-type stackings with the additional weak interspersed features suggesting minor contributions from
rhombohedral-type stackings.  The dominance of Bernal-type stackings for a quadlayer contrasts with the coexistence of Bernal and
rhombohedral stackings in the trilayer, and suggests the role of the second nearest neighbor in stabilizing Bernal stacking in bulk
graphite.

\begin{table}
\caption{\label{tab:table}TB band parameters to reproduce measured bandstructure for $N$=1-4 layers graphene (present work) and
$N$=$\infty$ (graphite, Ref. \cite{Zho06}). The electron density $n$ is measured in $10^{-3}$ electron per 2D unit cell. 
}

\def\vunits{($10^{6} \frac{m}{s}$)}
\begin{ruledtabular}
\begin{tabular}{ccc|cccccc}
$N$      &$v$           & $n$ & \ED     &$E_{1}$ & $E_{2}$ & $E_{3}$ & $E_{4}$ &$\gamma_{1}$  \\
         &\vunits       &     & \multicolumn{6}{c}{(eV)}                            \\
\hline
1	 & 1.10		& 7.4 & -0.44   &-0.44   & 	   &	     &         &              \\
2	 & 1.05		& 8.2 & -0.30   &-0.36   & -0.24   &	     &         & 0.48         \\
3	 & 1.02		& 8.6 & -0.21   &-0.31   & -0.17   & -0.15   &         & 0.48         \\
4	 & 1.02		& 7.4 & -0.15   &-0.27   & -0.13   & -0.10   & -0.10   & 0.45         \\
$\infty$ & 0.91	        &     &         &        & 	   &	     &         & $\sim 0.35$         \\
\end{tabular}
\end{ruledtabular}
\end{table}

The tight binding parameters (Table I) reveal important details of the interlayer coupling and screening of graphene.  Notice that
our interlayer hopping integrals for multilayer graphene are significantly larger than that of bulk graphite, around 0.35 eV
\cite{Zho06, Cha91}.  In bilayer graphene, the interlayer hopping integral is reported to increase upon increasing carrier
concentration \cite{Oht06}.  We postulate that the larger interlayer hopping integral is similarly brought about by the higher
carrier concentrations than that of graphite.

The change in the out-of-plane coupling of the electronic states is associated with the transition from pure-2D to quasi-2D as the
number of layers increases towards eventual 3D (bulk) character.  We study this transition by changing the electron wave vector normal to the
graphene planes through a change of photon energy $h\nu$.  Fig.\ \ref{fig:piBS}(e-h) shows the normalized photoemission intensities of \PI\
bands at \EF$-1$ eV for 1-4 layers graphene as a function of electron momentum parallel (\kpar) and normal (\kperp) to the surface,
corresponding to $h\nu$ $\sim$ 77 - 210 eV. The maps in Fig.\ \ref{fig:piBS}(e-h) were normalized by dividing out the total
intensity between \EF\ and \EF$-1.5$ eV along M-K-$\Gamma$ direction (Fig.\ \ref{fig:piBS}(a-d)) at each $h\nu$.  It is assumed that \kparnum\ at the K point with the same inner potential
($\sim$16.5 eV) as graphite \cite{Zho06b,Oht06b}.

As expected for 2D systems, the \PI\ bands display no dispersion along \kperp\ (Fig.\ \ref{fig:piBS}(e-h)).  For single layer
graphene, the photoemission cross section decays smoothly and monotonically (this slow variation is normalized out in the data Fig.\
\ref{fig:piBS}(e)).  For 2 to 4 layer graphene (Fig.\ \ref{fig:piBS}(f-h)), however, the photoemission cross sections oscillate with
a periodicity of $\sim$2.0~\AA$^{-1}$.  This periodicity does not match with the reciprocal lattice vector of bulk graphite
(0.932~\AA$^{-1}$), which includes two graphene layers in the unit cell, but is close to the reciprocal interlayer distance of graphite
(1.86~\AA$^{-1}$).  The slightly longer periodicity of the observed intensity oscillation may also imply a reduced interlayer
distance in our films, although precise assessment of the \kperp\ periodicity requires careful estimation of the inner potential.  A
reduction in the spacing for bilayer graphene was previously attributed to the increased screening with carrier
concentration\cite{Oht06}.

The intensity oscillations with \kperp\ for N$\geq2$ are similar to oscillations reported for quantized thin films states
\cite{Han97, Mug00}, since the increase in the number of the features for 2 to 4 layers can be viewed as resulting from the
quantum-well like nature of the wave functions in the graphene layers.  The lack of such oscillations for monolayer graphene results
not just from the trivial lack of overlying graphene layers, but also due to a notable lack of coupling to the substrate, consistent with the lack of \PI\ bands in the underlying SiC interface layer \cite{Emt06}.  This is in contrast to surface states \cite{Louie80, Hof02, Lob06}, in which the photoemission cross section has a maximum near vertical transitions of the bulk crystal, suggesting that single layer graphene wave functions have nearly ideal 2D character.  

\begin{figure}
\begin{center}
    \includegraphics[width=8.4cm]{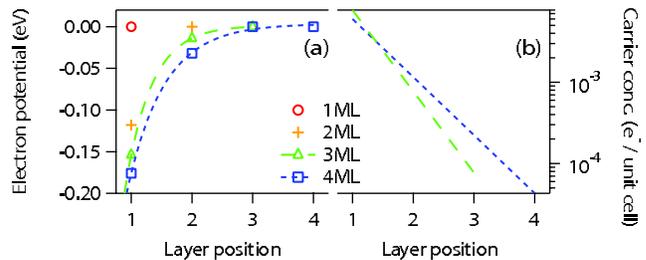}
% f02003_40048_92_01649_40224ML.pxp \vspace{1mm}
\caption{\label{fig:poten} (color online) Potential and carrier concentration profiles of the multilayer graphene as a function of the layer
positions.  The electron potentials are shifted in the way that the potential of the outermost graphene layer is at zero.}
\end{center}
\end{figure}

The extracted on-site Coulomb potentials (Table I) may be used to estimate the screening length and the distribution of carrier
concentrations.  Fitting the on-site potentials (plotted in (Fig.\ \ref{fig:poten}(a) with the outermost layers' potentials aligned
to the zero-reference level) to a simple exponential decay allows direct determination of the interlayer screening lengths
for three and four layer graphene, 1.45~\AA\ and 2.06~\AA, respectively.  This trend of longer screening length for thicker films
follows naturally from the decreased charge per layer, resulting in weaker interlayer screening.  In turn, the estimated screening
lengths are smaller than the reported value for graphite (3.8-5~\AA\ \cite{Pie78, DiV84}), which has a much smaller carrier
concentration than our films.

The total charge density $n$ was determined by measuring \PI-band Fermi surface areas, and was found to be almost invariant with
film thickness or stacking order (Table I); this explains the observed decrease of (\EF$-$\ED) with film thickness $N$ (Fig.\ 2a-d).  The $\sim$10
$\%$ deviation of the carrier concentration for different thicknesses is due to the accuracy of the fitting procedure.

Using Poisson's equation and the exponentially fitted profiles (Fig.\ \ref{fig:poten}(a)), the distribution of the measured total charge
density $n$ (Table I) across the layers could be estimated for $N$=3 and 4 graphene layers (Fig.\ \ref{fig:poten}(b)).  The carrier
concentration decreases by about one order of magnitude for each adjacent layer.  For the reported multilayer graphene devices
\cite{Mor05}, the screening length is expected to be larger than the present case because of lower carrier concentration, therefore
the carrier concentration profile is expected to be less steep.  This suggests that the carriers in less-doped multilayers are
distributed across several layers.

While the electron potentials of a laterally confined surface state of vicinal Au(111) were extracted using a combination of ARPES
and scanning tunneling spectroscopy \cite{Mug03, Did06}, the evaluation of the potential, screening length, and carrier
concentrations solely from ARPES band structure measurements is unique to the present study. The present analysis is only
possible because the topology of the \PI\ states is very sensitive to the on-site Coulomb potential, and because of the very
high energy and momentum resolutions of the experiment.

In summary, we identify the \PI\ band splitting due to interlayer interactions and different stacking sequences of the graphene
layers.  The interlayer interaction alters the character of the \PI\  wave function from pure-2D in a single layer to quasi-2D in
multilayer graphene.  Exploiting the sensitivity of the \PI\ states to the electron potential, the profiles of potential and
carrier concentration in each layer and the screening length are estimated for three and four layer graphene.  We find the
interlayer hopping integral and screening length  deviate significantly from those of graphite because of the altered carrier
concentrations, which illustrate the unique electronic properties of graphene layers.  

The authors thank K.V. Emtsev, J. M. Carlsson, F. Guinea and A. Mascaraque for fruitful discussions.  We are grateful to K. V.
Emtsev and S. Reshanov for hydrogen treatment of SiC. This work and the Advanced Light Source were supported
by the Director, Office of Science, Office of Basic Energy Sciences, of the U.S. Department of Energy under Contract No.\
DE-AC02-05CH11231.  T.O. and K.H. were supported by the Max Planck Society and the European Science Foundation under the EUROCORES
SONS program.

%\bibliography{FLG}

\end{document}